\documentclass[11pt]{article}
\usepackage{amsmath,amssymb,amsthm,amsxtra,overpic,bbm,bm,epsfig,subfigure}
\usepackage{color}
\usepackage{slashed,stmaryrd}

\def\thefootnote{\fnsymbol{footnote}}

\begin{document}

\vspace{0.2cm}

\begin{center}
{\Large\bf Relic Right-handed Dirac Neutrinos and Implications for Detection of Cosmic Neutrino Background}
\end{center}

\vspace{0.2cm}

\begin{center}
{\bf Jue Zhang $^{a}$} \footnote{E-mail: zhangjue@ihep.ac.cn}
\quad {\bf Shun Zhou $^{a,~b}$} \footnote{E-mail: zhoush@ihep.ac.cn}
\\
{$^a$Institute of High Energy Physics, Chinese Academy of
Sciences, Beijing 100049, China \\
$^b$Center for High Energy Physics, Peking University, Beijing 100871, China}
\end{center}

\vspace{1.5cm}
\begin{abstract}
It remains to be determined experimentally if massive neutrinos are Majorana or Dirac particles. In this connection, it has been recently suggested that the detection of cosmic neutrino background of left-handed neutrinos $\nu^{}_{\rm L}$ and right-handed antineutrinos $\overline{\nu}^{}_{\rm R}$ in future experiments of neutrino capture on beta-decaying nuclei (e.g., $\nu^{}_e + {^3{\rm H}} \to {^3}{\rm He} + e^-$ for the PTOLEMY experiment) is likely to distinguish between Majorana and Dirac neutrinos, since the capture rate is twice larger in the former case. In this paper, we investigate the possible impact of right-handed neutrinos on the capture rate, assuming that massive neutrinos are Dirac particles and both right-handed neutrinos $\nu^{}_{\rm R}$ and left-handed antineutrinos $\overline{\nu}^{}_{\rm L}$ can be efficiently produced in the early Universe. It turns out that the capture rate can be enhanced at most by $28\%$ due to the presence of relic $\nu^{}_{\rm R}$ and $\overline{\nu}^{}_{\rm L}$ with a total number density of $95~{\rm cm}^{-3}$, which should be compared to the number density $336~{\rm cm}^{-3}$ of cosmic neutrino background. The enhancement has actually been limited by the latest cosmological and astrophysical bounds on the effective number of neutrino generations $N^{}_{\rm eff} = 3.14^{+0.44}_{-0.43}$ at the $95\%$ confidence level. For illustration, two possible scenarios have been proposed for thermal production of right-handed neutrinos in the early Universe.
\end{abstract}

\begin{flushleft}
\hspace{0.8cm} PACS number(s): 14.60.St, 95.35.+d, 98.70.Vc
\end{flushleft}

\def\thefootnote{\arabic{footnote}}
\setcounter{footnote}{0}

\newpage

\section{Introduction}

Although a number of elegant neutrino oscillation experiments in the past few decades have well established that neutrinos are massive particles, it is still unclear whether massive neutrinos are of Dirac or Majorana nature~\cite{Agashe:2014kda,Wang:2015rma}. Thus far, tremendous efforts have been placed on the experimental searches for neutrinoless double-beta ($0\nu\beta\beta$) decays, which take place only if lepton number violation exists and massive neutrinos are Majorana particles~\cite{Pas:2015eia,Bilenky:2014uka,Bilenky:2012qi,Rodejohann:2011mu}. The experimental discovery of $0\nu\beta\beta$ decays will provide us with a robust evidence for Majorana neutrinos. However, in case that $0\nu\beta\beta$ decays are not detected in all the future $0\nu\beta\beta$ experiments, it is still possible that neutrinos are Majorana particles, if neutrino mass ordering is normal (i.e., $m^{}_1 < m^{}_2 < m^{}_3$) and an intricate cancellation occurs in the effective neutrino mass relevant for $0\nu\beta \beta$ decays (see, e.g., Ref.~\cite{Xing:2015zha}). In this case, another independent approach should be utilized to probe the Dirac or Majorana nature of massive neutrinos.

More than fifty years ago, Weinberg pointed out~\cite{Weinberg:1962zza} that the cosmic neutrino background (C$\nu$B) predicted by the standard Big Bang theory of cosmology can be detected via neutrino capture on beta-decaying nuclei, e.g., $\nu^{}_e + {^{3}{\rm H}} \to {^{3}{\rm He}} + e^-$. This possibility has been extensively studied in many recent works~\cite{Cocco:2007za,Lazauskas:2007da,Blennow:2008fh,Li:2010sn,Faessler:2011qj,Long:2014zva}. In particular, for the future experiment PTOLEMY~\cite{Betts:2013uya} with 100 grams of tritium, the capture rate $\Gamma(\nu^{}_e + {^{3}{\rm H}} \to {^{3}{\rm He}} + e^-)$ has been found to be~\cite{Long:2014zva}
\begin{eqnarray} \label{eq:rates}
\Gamma^{}_{\rm M} \approx 8~{\rm yr}^{-1} & ({\rm Majorana}) \; ; \hspace{1cm}
\Gamma^{}_{\rm D} \approx 4~{\rm yr}^{-1} & ({\rm Dirac}) \; .
\end{eqnarray}
These results have profound implications for cosmology and elementary particle physics. First, a successful detection of C$\nu$B is very important to further verify the standard theory of cosmology~\cite{Kolb:1990vq,Dodelson:2003ft,Weinberg:2008zzc}, and serves as a unique way to probe our Universe back to the time when it was just one second old. We already have an excellent example that the precise measurements of cosmic microwave background (CMB) have given valuable information on the Universe at the age of $3.8\times 10^5$ years, and greatly improved our knowledge on the cosmology. Second, the relation $\Gamma^{}_{\rm M} = 2 \Gamma^{}_{\rm D}$ between the capture rates in Eq.~(\ref{eq:rates}) offers a novel way to distinguish between Dirac and Majorana neutrinos. In this paper, we concentrate on the second point and take it more seriously.

It is worthwhile to emphasize that the contributions from right-handed components of massive Dirac neutrinos are completely neglected in the calculations leading to Eq.~(\ref{eq:rates}). See, Ref.~\cite{Long:2014zva}, for more details. An immediate question is how the right-handed Dirac neutrinos are produced in our Universe, in the standard theories of particle physics and cosmology, and whether their abundance can be safely neglected. The second question is how the right-handed Dirac neutrinos affect the detection of C$\nu$B, i.e., the capture rate in Eq.~(\ref{eq:rates}), if they are copiously generated in the early Universe and survive today as a cosmic background. In order to answer these two questions, we assume that massive neutrinos are Dirac particles, and investigate carefully their production and evolution in the early Universe, both within and beyond the standard model of particle physics (SM).

The remaining part of the present paper is organized as follows. In Sec. 2, the thermal production of right-handed neutrinos in the minimal extension of the SM with massive Dirac neutrinos is reviewed. The production rate turns out to be extremely small and can be neglected. Then, we investigate the cosmological constraint on the relic density of right-handed neutrinos in Sec. 3, assuming that they can be thermalized in the early Universe in the scenarios beyond the SM. Subsequently, in Sec. 4, two possible scenarios have been presented to show that they can indeed be thermally produced if the primordial magnetic fields or secret interactions among right-handed neutrinos exist. Sec. 5 is devoted to the impact of relic right-handed neutrinos on the detection of C$\nu$B. Finally, we summarize our main results in Sec. 6.

\section{The Extended SM}

We first briefly review the minimal extension of the SM with three right-handed neutrinos $\nu^{}_{\alpha \rm R}$ (for $\alpha = e, \mu, \tau$), which are singlets under the ${\rm SU}(2)^{}_{\rm L} \times {\rm U}(1)^{}_{\rm Y}$ gauge symmetry. The relevant Lagrangian reads
\begin{eqnarray}\label{eq:lag}
{\cal L} = {\cal L}^{}_{\rm SM} + \overline{\nu^{}_{\alpha \rm R}} i\slashed{\partial}\nu^{}_{\alpha \rm R} - \left[ \overline{\ell^{}_{\alpha {\rm L}}} \left(Y^{}_\nu\right)^{}_{\alpha \beta} \tilde{H} \nu^{}_{\beta {\rm R}} + {\rm h.c.}\right] \; ,
\end{eqnarray}
where ${\cal L}^{}_{\rm SM}$ stands for the SM Lagrangian, $\ell^{}_{\alpha {\rm L}} \equiv \left(\nu^{}_{\alpha {\rm L}}, l^{}_{\alpha {\rm L}}\right)^{\rm T}$ and $\tilde{H} = {\rm i} \sigma^{}_2 H^*$ denote respectively lepton and Higgs doublets, $\left(Y^{}_\nu\right)^{}_{\alpha \beta}$ for $\alpha, \beta = e, \mu, \tau$ are Dirac neutrino Yukawa couplings. After the Higgs field acquires its vacuum expectation value $\langle H \rangle \equiv v \approx 174~{\rm GeV}$ and the gauge symmetry is spontaneously broken down, one obtains the Dirac neutrino mass matrix $M^{}_{\rm D} = Y^{}_\nu v$, which can be diagonalized via a bi-unitary transformation $U^\dagger_{\rm L} M^{}_{\rm D} U^{}_{\rm R} = {\rm diag}\{m^{}_1, m^{}_2, m^{}_3\}$. In the mass basis, $\nu^{}_{i {\rm L}}$ and $\nu^{}_{i {\rm R}}$ constitute a massive Dirac spinor $\nu^{}_i = \nu^{}_{i {\rm L}} + \nu^{}_{i {\rm R}}$. In the following, we refer to this minimal extension of the SM with massive Dirac neutrinos as the extended SM. As indicated by the latest Planck results on CMB~\cite{Ade:2015xua}, the sum of neutrino masses $\Sigma = m^{}_1 + m^{}_2 + m^{}_3$ is strictly constrained, i.e., $\Sigma < 0.23~{\rm eV}$ at the $95\%$ confidence level (C.L.). Therefore, for Dirac neutrinos, we are left with two serious problems. First, a global ${\rm U}(1)$ symmetry corresponding to the lepton number conservation has to be imposed on the generic Lagrangian in Eq.~(\ref{eq:lag}) in order to forbid a Majorana mass term $\overline{\nu^{\rm C}_{\rm R}} M^{}_{\rm R} \nu^{}_{\rm R}$, which otherwise is allowed by the SM gauge symmetry. Second, the neutrino Yukawa coupling constants $y^{}_i \equiv m^{}_i/v \sim 10^{-12}$ in the mass basis are smaller by twelve orders of magnitude than the top-quark Yukawa coupling $y^{}_t \sim {\cal O}(1)$. This exaggerates the fermion mass hierarchy problem of the SM. In a realistic model of Dirac neutrinos, these two problems should be properly addressed. However, we temporarily put them aside and focus on the cosmological implications in the following discussions.

After specifying the theoretical framework, we are now in a position to consider the production of $\nu^{}_{\rm R}$ in the early Universe. In fact, this task has already been accomplished in Ref.~\cite{Antonelli:1981eg}. However, it is instructive to revisit this problem in view of recent progress in neutrino physics (e.g., the establishment of massive neutrinos) and the discovery of Higgs boson (e.g., the observation of Higgs-fermion interactions). Following the notations in Ref.~\cite{Kolb:1990vq}, we can calculate the number density $n^{}_a$ of a particle species $a$, which is involved in the interaction $X \leftrightarrow a + Y$, via the Boltzmann equation
\begin{eqnarray} \label{eq:Boltzmann}
\frac{{\rm d}n^{}_a}{{\rm d}t} + 3 H n^{}_a = - \sum_{X \leftrightarrow a + Y} \left[ \frac{n^{}_a}{n^{\rm eq}_a} \frac{n^{}_Y}{n^{\rm eq}_Y} \gamma(a+Y \to X) - \frac{n^{}_X}{n^{\rm eq}_X} \gamma(X \to a + Y)\right] \; ,
\end{eqnarray}
where $n^{\rm eq}_i$ (for $i = a, X, Y$) are the number densities in thermal equilibrium, and $H$ is the Hubble expansion rate. Those two terms in the parentheses on the right-hand side of Eq.~(\ref{eq:Boltzmann}) stand for the absorption and production rates, respectively. More explicitly, the collision term is given by
\begin{eqnarray} \label{eq:gamma}
\gamma(X \to a + Y) = \int \frac{{\rm d}^3{\bf p}^{}_X}{(2\pi)^3 2p^0_X} \frac{{\rm d}^3 {\bf p}^{}_a}{(2\pi)^3 2p^0_a} \frac{{\rm d}^3 {\bf p}^{}_Y}{(2\pi)^3 2p^0_Y} (2\pi)^4 \delta^4(p^{}_X - p^{}_a - p^{}_Y) e^{-p^0_X/T} \left|{\cal M}(X \to a + Y)\right|^2 \; ,
\end{eqnarray}
where $p^{}_i$ (for $i = a, X, Y$) are the four-momenta, $T$ is the temperature, and $|{\cal M}(X \to a + Y)|^2$ should be summed but not averaged over the internal degrees of freedom of the initial and final states. In general, $X$ and $Y$ can also be a set of multiple-particle states.

In the early Universe, when the temperature is extremely high $T \gg T^{}_{\rm EW}$ with $T^{}_{\rm EW} \approx 200~{\rm GeV}$ being the critical temperature for electroweak phase transition, the SM gauge symmetry is restored and all the SM particles are massless, except for the Higgs boson. The right-handed neutrinos $\nu^{}_{i{\rm R}}$ only experience the Yukawa interactions, as given in Eq.~(\ref{eq:lag}), so the production and absorption of $\nu^{}_{i{\rm R}}$ are governed by the tiny Yukawa couplings $y^{}_i$. Note that we are working in the mass basis in the sense that the Yukawa coupling matrix is diagonal. In this case, the dominant processes for $\nu^{}_{i{\rm R}}$ production should be Higgs boson decays $H \to \overline{\nu^{}_{i {\rm L}}} + \nu^{}_{i {\rm R}}$, top-quark scattering $\overline{t^{}_{\rm L}} + t^{}_{\rm R} \to H \to \overline{\nu^{}_{i {\rm L}}} + \nu^{}_{i {\rm R}}$ and gauge boson scattering $V + V \to H \to \overline{\nu^{}_{i {\rm L}}} + \nu^{}_{i {\rm R}}$. Taking only the decays and inverse decays into account, we can immediately figure out the decay rate $\Gamma^{}_H = y^2_i M^{}_H/(32\pi)$, and thus the corresponding collision term
\begin{eqnarray} \label{eq:gammaH}
\gamma(H \to \overline{\nu^{}_{i {\rm L}}} \nu^{}_{i {\rm R}}) = \frac{M^{}_H \Gamma^{}_H T^2}{2\pi^2} K^{}_1(M^{}_H/T) \equiv \gamma^{}_{\rm D}\; ,
\end{eqnarray}
where $K^{}_n$ is the $n$-th order modified Bessel function. The contributions from the scattering processes are on the same order as that from decays. Because of finite-temperature effects~\cite{Weldon:1982bn}, the gauge interactions result in thermal lepton masses $M^2_{\ell}(T) = (3g^2 + {g^\prime}^2) T^2/32$, implying a slight reduction of the decay rate compared to the result at zero temperature. Moreover, both gauge and top-quark Yukawa interactions give rise to a thermal Higgs mass $M^2_H(T) = (8M^2_W + M^2_Z + 2m^2_t + M^2_H)(T^2 - T^2_{\rm EW})/(8v^2)$, where the gauge boson, top-quark and Higgs boson masses are evaluated at $T=0$. Given the latest values $m^{}_t = 173~{\rm GeV}$ and $M^{}_H = 125~{\rm GeV}$, we can obtain $M^{}_H(T)/T \approx 3/4$ in the limit of $T \gg T^{}_{\rm EW}$. As a consequence, the Higgs mass $M^{}_H$ in Eq.~(\ref{eq:gammaH}) should be replaced by the thermal one $M^{}_H(T) \approx 3T/4$. Both Higgs bosons and left-handed neutrinos are well in thermal equilibrium due to the efficient gauge interactions, so Eq.~(\ref{eq:Boltzmann}) can be simplified to
\begin{eqnarray} \label{eq:simple}
\frac{{\rm d}n^{}_{\nu^{}_{i{\rm R}}}}{{\rm d}t} + 3 H n^{}_{\nu^{}_{i{\rm R}}} = \left(1 - \frac{n^{}_{\nu^{}_{i{\rm R}}}}{n^{\rm eq}_{\nu^{}_{i{\rm R}}}}\right) \gamma^{}_{\rm D} \; ,
\end{eqnarray}
where the in-equilibrium density $n^{\rm eq}_{\nu^{}_{i{\rm R}}} = T^3/\pi^2$ for the Boltzmann distribution with a vanishing chemical potential. For comparison, the number density of photons in thermal equilibrium is given by $n^{\rm eq}_\gamma = 2 n^{\rm eq}_{\nu^{}_{i{\rm R}}} = 2T^3/\pi^2$. Now it is evident from Eq.~(\ref{eq:simple}) that if the creation rate $\Gamma^{}_{\nu^{}_{i{\rm R}}} \equiv \gamma^{}_{\rm D}/n^{\rm eq}_{\nu^{}_{i{\rm R}}}$ is much smaller than the Hubble expansion rate $H = 1.66 \sqrt{g^{}_*} T^2/M^{}_{\rm pl}$, where $g^{}_*$ is the number of relativistic degrees of freedom and $M^{}_{\rm pl} = 1.2\times 10^{19}~{\rm GeV}$ is the Planck mass scale, the production of $\nu^{}_{i{\rm R}}$ will be inefficient. With the help of Eq.~(\ref{eq:gammaH}), it is straightforward to derive
\begin{eqnarray}
R \equiv \Gamma^{}_{\nu^{}_{i{\rm R}}}/H \approx 10^{-3} \frac{y^2_i}{\sqrt{g^{}_\ast}} \frac{M^{}_{\rm pl}}{T} \; ,
\end{eqnarray}
which is valid for $T \gg T^{}_{\rm EW}$. For $y^{}_i \approx 10^{-12}$ and $g^{}_\ast = 106.75$ in the SM at $T = 10^7~{\rm GeV}$, one arrives at $R \approx 10^{-16}$, which is larger by several orders of magnitude than the result in Ref.~\cite{Antonelli:1981eg}, where only neutral-current interactions are included. For $T = 10^3~{\rm GeV}$, we have an even larger ratio $R \sim 10^{-12}$. However, such a small production rate indicates that $\nu^{}_{i{\rm R}}$ will never be populated in the early Universe, which should hold as well at any temperature above $T^{}_{\rm EW}$.

Below the electroweak phase transition, i.e., $T \ll T^{}_{\rm EW}$, the Higgs bosons, top quarks and weak gauge bosons have already decayed away, and the dominant production channel of $\nu^{}_{i{\rm R}}$ is through the conversion from $\nu^{}_{i{\rm L}}$ due to the presence of Dirac masses~\cite{Antonelli:1981eg}. Given the production rate of $\nu^{}_{i{\rm L}}$ being $\Gamma^{}_{\nu^{}_{i{\rm L}}} \sim G_{\rm F}^2 T^5$, where $G_{\rm F}$ is the Fermi coupling constant, we obtain $\Gamma^{}_{\nu^{}_{i{\rm R}}} \sim G_{\rm F}^2 T^3 m_i^2$, and thus $R \sim 10^{-10}$ for $T = 10~{\rm GeV}$; or $R \sim 10^{-14}$ for $T = 1~{\rm MeV}$. Therefore, it becomes clear that $\nu^{}_{i{\rm R}}$ in the extended SM can never be abundantly produced in our Universe.

\section{Cosmological Constraints}

However, in some new physics scenarios beyond the SM, right-handed Dirac neutrinos can be copiously generated in the early Universe. Before going to any details of new physics models, we simply assume that $\nu^{}_{\rm R}$ can be thermally produced and then decouple from the plasma of SM particles at a freeze-out temperature $T^{\rm R}_{\rm fo}$, which needs not to be specified at this moment. With such a setup, our discussions will be applicable to more general cases. In addition to the C$\nu$B, $\nu^{}_{\rm R}$ will be a new kind of cosmic background, which is restrictively constrained by the cosmological observations, such as Big Bang Nucleosynthesis (BBN) and CMB.

At the freeze-out temperature, where the decoupling is taken to be instantaneous for simplicity, the number densities of $\nu^{}_{\rm R}$ and $\nu^{}_{\rm L}$ are equal, i.e., $n_{\nu_{\rm R}^{}}^{}(T_{\rm fo}^{\rm R}) =  n_{\nu_{\rm L}^{}}^{} (T_{\rm fo}^{\rm R})$. In the later evolution of the Universe, $\nu^{}_{\rm L}$ remains in good contact with the other SM particles via gauge interactions, and its number density is determined by the Fermi-Dirac distribution with a temperature $T$. However, $\nu_{\rm R}^{}$ gets diluted by the expansion, and its number density at temperature $T$ is given by
\begin{eqnarray}
n_{\nu_{\rm R}^{}}^{} (T) = n_{\nu_{\rm R}^{}}^{}(T_{\rm fo}^{\rm R}) \left[ \frac{a(T_{\rm fo}^{\rm R})}{a(T)} \right]^3,
\end{eqnarray}
where $a(T)$ is the scale factor as a function of the temperature $T$. On the other hand, assuming adiabatic expansion of the Universe, we find that the conservation of entropy in the thermal bath leads to
\begin{eqnarray} \label{eq:entropy_conservation}
\frac{ g_{*s}^{}(T) T^3 }{g_{*s}^{}(T_{\rm fo}^{\rm R}) (T_{\rm fo}^{\rm R})^3} =  \left[ \frac{a(T_{\rm fo}^{\rm R})}{a(T)} \right]^3,
\end{eqnarray}
where $g_{*s}^{}$ stands for the number of effective degrees of freedom. Note that we have $g^{}_{*s}(T) = g^{}_*(T)$ before the neutrino decoupling. Since the left-handed neutrino $\nu_{\rm L}^{}$ is relativistic and in thermal equilibrium before it decouples when $T \gtrsim 1$ MeV, its number density scales as $n_{\nu_{\rm L}^{}} \propto T^3$. Therefore, at the temperature $T_{\rm fo}^{\rm L}$ when $\nu_{\rm L}^{}$ begins to freeze out, the ratio of the number densities of $\nu_{\rm R}^{}$ and $\nu_{\rm L}^{}$ reads
\begin{eqnarray}
\frac{n_{\nu_{\rm R}^{}}^{}(T_{\rm fo}^{\rm L})}{n_{\nu_{\rm L}^{}}^{}(T_{\rm fo}^{\rm L})} = \frac{g_{*s}^{}(T_{\rm fo}^{\rm L})}{g_{*s}^{}(T_{\rm fo}^{\rm R})},
\end{eqnarray}
where the relations $n^{}_{\nu^{}_{\rm L}}(T^{\rm L}_{\rm fo})/n^{}_{\nu^{}_{\rm L}}(T^{\rm R}_{\rm fo}) = (T^{\rm L}_{\rm fo}/T^{\rm R}_{\rm fo})^3$ and $n^{}_{\nu^{}_{\rm L}}(T^{\rm R}_{\rm fo}) = n^{}_{\nu^{}_{\rm R}}(T^{\rm R}_{\rm fo})$ have been used. The ratio of right-handed and left-handed neutrino number densities is fixed by $g^{}_{*s}$ at the freeze-out temperature of $\nu^{}_{\rm R}$, since we know $T^{\rm L}_{\rm fo} \approx 1~{\rm MeV}$ and $g^{}_{*s}(T^{\rm L}_{\rm fo}) = 10.75$. If $T^{\rm R}_{\rm fo} > T^{}_{\rm EW}$, we have $g^{}_{*s}(T^{\rm R}_{\rm fo}) = 106.75$, and thus the right-to-left ratio of neutrino number densities is $n^{}_{\nu^{}_{\rm R}}/n^{}_{\nu^{}_{\rm L}} \approx 0.1$. Since $T < T_{\rm fo}^{\rm L}$, both $\nu_{\rm L}^{}$ and $\nu_{\rm R}^{}$ are freely streaming, the above ratio is then unchanged at the present time. As is well known, the C$\nu$B consists of left-handed neutrinos $\nu^{}_{\rm L}$ and right-handed antineutrinos $\overline{\nu}^{}_{\rm R}$, whose number density is $336~{\rm cm}^{-3}$. Now we also have right-handed neutrinos $\nu^{}_{\rm R}$ and left-handed antineutrinos $\overline{\nu}^{}_{\rm L}$ as a cosmic background with a total number density $34~{\rm cm}^{-3}$. Even larger number densities of $\nu^{}_{\rm R}$ and $\overline{\nu}^{}_{\rm L}$ are also possible for a lower freeze-out temperature, as we shall show later.

%%%%%%%%%%%%%%%%%%%%%%%%%% Fig. 1 %%%%%%%%%%%%%%%%%%%%%%%%%%%%
\begin{figure}[t!]
\begin{center}
\subfigure{
\includegraphics[width=0.8\textwidth]{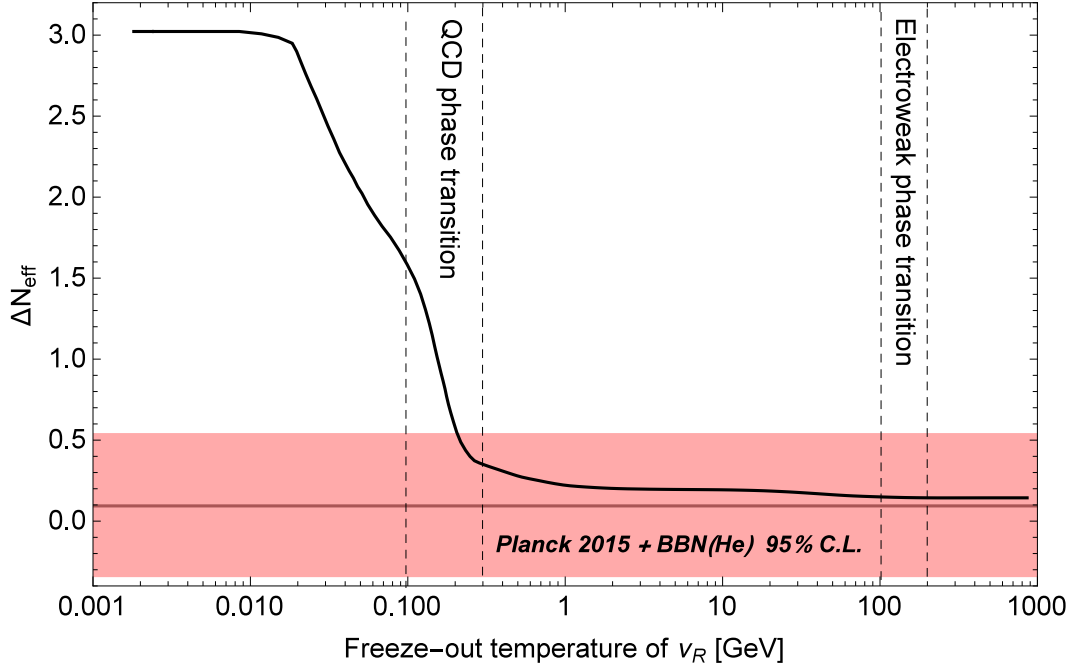} }
\vspace{-0.1cm}
\caption{The extra effective number of neutrino generations $\Delta N^{}_{\rm eff} \equiv N^{}_{\rm eff} - 3.046$ during the era of Big Bang Nucleosynthesis has been calculated by choosing different values of the freeze-out temperature of right-handed neutrinos $T^{\rm R}_{\rm fo}$. The shaded area represents current cosmological and astrophysical constraints $\Delta N^{}_{\rm eff} = 0.10^{+0.44}_{-0.43}$ at the $95\%$ C.L., where the {\it Planck} TT + lowP + BAO data sets~\cite{Ade:2015xua} are combined with the helium abundance measurements~\cite{Aver:2013wba}. The number of relativistic degrees of freedom $g^{}_{*s}(T)$ has been taken from Refs.~\cite{Steigman:2012nb,Laine:2006cp}.} \label{fig:DeltaN}
\end{center}
\end{figure}
%%%%%%%%%%%%%%%%%%%%%%%%%%%%%%%%%%%%%%%%%%%%%%%%%%%%%%%%%%%%%%

The presence of extra radiation, such as $\nu^{}_{\rm R}$ under consideration, could substantially modify the predictions for primordial abundances of light nuclear elements by the standard BBN theory and the power spectrum of CMB. As usual, the contribution of extra relativistic particles $x$ to the total energy density of radiation at the CMB temperature is parameterized as
\begin{eqnarray} \label{eq:definition}
\rho^{}_{\rm r} = \rho^{}_\gamma + \rho^{}_x = \left[1 + N^{}_{\rm eff} \frac{7}{8} \left(\frac{4}{11}\right)^{4/3}\right]\rho^{}_\gamma \; ,
\end{eqnarray}
where $N^{}_{\rm eff} = 3.046$ is expected if only three generations of neutrinos in the SM exist. That the value is not exactly three can be ascribed to the non-thermal distortion of neutrino energy spectra, as neutrinos are still slightly coupled to the thermal bath when electrons and positrons annihilate into photons~\cite{Mangano:2005cc,Iocco:2008va}. Now the contribution from $\nu^{}_{\rm R}$ should also be taken into account. In a similar way, one can derive the relationship between the energy densities of $\nu^{}_{\rm R}$ and $\nu^{}_{\rm L}$ at the CMB temperature $T^{}_{\rm CMB} \approx 0.3~{\rm eV}$ as follows
\begin{eqnarray}
\frac{\rho^{}_{\nu^{}_{\rm R}}(T^{}_{\rm CMB})}{\rho^{}_{\nu^{}_{\rm L}}(T^{}_{\rm CMB})} = \left[\frac{g^{}_{*s}(T^{}_{\rm CMB})}{g^{}_{*s}(T^{R}_{\rm fo})}\right]^{4/3} \left( \frac{11}{4} \right)^{4/3} \frac{N^\nu_{}}{N^{\nu_{\rm L}^{}}_{\rm eff}}\; ,
\end{eqnarray}
where $N^\nu_{} = 3$ because of three generations of neutrinos, and $N^{\nu_{\rm L}^{}}_{\rm eff} = 3.046$ originating from the left-handed neutrinos $\nu_{\rm L}^{}$. The extra $N^{}_{\rm eff}$ due to the presence of right-handed neutrinos $\nu^{}_{\rm R}$ can then be defined as $\Delta N^{}_{\rm eff} \equiv N^{}_{\rm eff} - N^{\nu_{\rm L}^{}}_{\rm eff}$, and it is found to be
\begin{eqnarray}
\Delta N^{}_{\rm eff} = \left[\frac{g^{}_{*s}(T^{}_{\rm CMB})}{g^{}_{*s}(T^{R}_{\rm fo})}\right]^{4/3} \left( \frac{11}{4} \right)^{4/3} N^\nu_{} \; .
\end{eqnarray}
Assuming that the left-handed neutrinos $\nu^{}_{\rm L}$ decouple instantaneously, the above $\Delta N^{}_{\rm eff}$ reduces to
\begin{eqnarray}
\Delta N^{}_{\rm eff} = \left[\frac{g^{}_{*s}(T_{\rm fo}^{\rm L})}{g^{}_{*s}(T^{R}_{\rm fo})}\right]^{4/3}  N^\nu_{} \; ,
\end{eqnarray}
in agreement with the results given in Refs.~\cite{Anchordoqui:2011nh,Anchordoqui:2012qu,SolagurenBeascoa:2012cz}.\footnote{To account for the non-instantaneous decoupling effects of $\nu^{}_{\rm L}$, we adopt the value of $g_{*s}(T^{}_{\rm CMB}) \approx 3.931$ in the following calculations, while its instantaneous decoupling limit should be 3.909.} According to the latest results from Planck Collaboration, the effective number of neutrino generations is determined to be~\cite{Ade:2015xua}
\begin{eqnarray} \label{eq:bound}
N^{}_{\rm eff} = 3.14^{+0.44}_{-0.43} \; , ~~ {\rm He} + {\it Planck} ~{\rm TT} + {\rm lowP} + {\rm BAO} \; ,
\end{eqnarray}
at the $95\%$ C.L. Any additional relativistic species present in the cosmic background will be stringently constrained by Eq.~(\ref{eq:bound}). In Fig.~\ref{fig:DeltaN}, we have calculated the extra number of neutrino species $\Delta N^{}_{\rm eff}$ by varying the freeze-out temperature $T^{\rm R}_{\rm fo}$. The cosmological bound is represented by the shaded area, while electroweak and QCD phase transitions are indicated by the dashed lines. In our calculations, the values of $g^{}_{*s}(T)$ have been taken from Refs.~\cite{Steigman:2012nb,Laine:2006cp}. Some comments are in order. First, the ranges of $T^{\rm R}_{\rm fo}$ below $200~{\rm MeV}$ are excluded by cosmological observations. This indicates the importance of QCD phase transition in diluting primordial relativistic particles. Second, for $T^{\rm R}_{\rm fo} \approx 200~{\rm MeV}$, the upper bound on $\Delta N^{}_{\rm eff}$ in Eq.~(\ref{eq:bound}) can be saturated, namely, $\Delta N^{}_{\rm eff} = 0.53$. In this situation, it is straightforward to get $n^{}_{\nu^{}_{\rm R}}/n^{}_{\nu^{}_{\rm L}} \approx 0.28$. Or equivalently, the number density of relic $\nu^{}_{\rm R}$ and $\overline{\nu}^{}_{\rm L}$  is $95~{\rm cm}^{-3}$. As we show later, these results affect significantly the detection of C$\nu$B.

\section{New Physics Scenarios}

Now we propose two possible scenarios to realize a thermal production of $\nu^{}_{\rm R}$ in the early Universe, and the freeze-out temperature should be above $T^{}_{\rm QCD} \approx 200~{\rm MeV}$ to evade cosmological bounds.

{\it Primordial Magnetic Fields.}---An important intrinsic property of massive Dirac neutrinos is that they can have nonzero magnetic dipole moments~\cite{Giunti:2014ixa,Giunti:2015gga}. If the SM is extended with massive Dirac neutrinos, one can obtain~\cite{Lee:1977tib,Fujikawa:1980yx}
\begin{eqnarray} \label{eq:munu}
\mu^{}_{\nu^{}_i} \approx 3\times 10^{-20} \left(\frac{m^{}_i}{0.1~{\rm eV}}\right) ~\mu^{}_{\rm B} \; ,
\end{eqnarray}
where $\mu^{}_{\rm B} \equiv e/2m^{}_e$ is the Bohr magneton. It is very likely that primordial magnetic fields exist in the early Universe, e.g., resulted from the electroweak phase transition~\cite{Vachaspati:1991nm,Baym:1995fk,Sigl:1996dm}, so the magnetic dipole interaction leads to an efficient production of $\nu^{}_{i{\rm R}}$ through the spin-flipping process $\nu^{}_{i{\rm L}} \to \nu^{}_{i{\rm R}}$~\cite{Enqvist:1994mb}. On the other hand, those primordial magnetic fields could also survive until today and serve as seed fields to explain the observed galactic magnetic fields around $B^{}_{\rm g} \approx 10^{-6}~{\rm G}$~\cite{Enqvist:1998fw}. Even though the magnetic fields can be generated during cosmological phase transitions, it remains unclear how the random magnetic field fluctuations are transformed into macroscopic-scale magnetic fields~\cite{Enqvist:1998fw}. For simplicity, we follow the phenomenological approach in Ref.~\cite{Enqvist:1994mb} and assume random magnetic fields with a scaling behavior as
\begin{eqnarray}
B(t, L) = B^{}_0 \left[\frac{a^{}_0}{a(t)}\right]^2 \left(\frac{L^{}_0}{L}\right)^p \; ,
\end{eqnarray}
where the term involving the scale factor $a(t)$ indicates the conservation of magnetic flux $B \sim a^{-2}$ during the expansion of the Universe. In addition, the index $p$ accounts for how the field strength depends on the physical spatial scale $L$. The initial domain size $L^{}_0$ and field strength $B^{}_0$ are determined by the production mechanism operated at the scale factor $a^{}_0$. In the radiation-dominated epoch, the relation $a(t) \propto t^{1/2} \propto T^{-1}$ holds, so one can convert the dependence of $B$ on the time into that on the temperature $T$. For the small-scale random magnetic fields $L^{}_{\rm W} \gg L^{}_0$, the $\nu^{}_{i{\rm L}} \to \nu^{}_{i{\rm R}}$ transition probability is~\cite{Enqvist:1994mb}
\begin{eqnarray}
\Gamma_{{\rm L}\to{\rm R}} = \frac{4}{3} \mu^2_{\nu^{}_i} B^2 L^{}_0 H^{-1} L^{-1}_{\rm W} \; ,
\end{eqnarray}
where $L^{-1}_{\rm W} \equiv \Gamma^{\rm tot}_{\rm W} \approx 30 G^2_{\rm F} T^5$ is the total weak interaction rate and the inverse of Hubble expansion rate $H^{-1}$ comes in as the largest time scale. For a suitable field strength $B^{}_0$ and domain size $L^{}_0$, there is no doubt that $\nu^{}_{i{\rm R}}$ can be brought into thermal equilibrium with $\nu^{}_{i{\rm L}}$. In order to ensure the decoupling of $\nu^{}_{i{\rm R}}$ at latest around $T^{}_{\rm QCD} \approx 200~{\rm MeV}$, we require the transition rate to be smaller than $H$ at $T^{}_{\rm QCD}$, namely,
\begin{eqnarray}
\mu^{}_{\nu^{}_i} B(T^{}_{\rm QCD}, l^{}_{\rm H}) \lesssim 6.7\times 10^{-3} \mu^{}_{\rm B}~{\rm G} \left(\frac{L^{}_{\rm W}}{L^{}_0}\right)^{1/2} \; ,
\end{eqnarray}
where $l^{}_{\rm H}(T^{}_{\rm QCD}) \equiv H^{-1}(T^{}_{\rm QCD}) \approx 3.5 \times 10^4~{\rm cm}$ and $L^{}_{\rm W}(T^{}_{\rm QCD}) \approx 1.6\times 10^{-2}~{\rm cm}$. Given the prediction of $\mu^{}_{\nu^{}_i}$ in Eq.~(\ref{eq:munu}), we can translate this constraint into an upper bound on the magnetic field $B^{}_0$ generated from the electroweak phase transition at $T^{}_{\rm EW}$, if the domain size fulfills $L^{}_0 \gtrsim L^{\rm min}_0(T) = 10^{-2}~{\rm cm} ~ ({\rm MeV}/T)$. Although the primordial magnetic fields in this case dissipate away at the time of BBN, it will not affect the production of $\nu^{}_{i{\rm R}}$ that is already completed at $T > T^{}_{\rm QCD}$. Taking $p = 1/2$ for example, we can get
\begin{eqnarray}
B^{}_0 \lesssim  10^{26}~{\rm G} \left(\frac{3\times 10^{-20} \mu^{}_{\rm B}}{\mu^{}_{\nu^{}_i}}\right) \; ,
\end{eqnarray}
where $L^{}_0 = L^{\rm min}_0(T^{}_{\rm QCD}) = 5\times 10^{-5}~{\rm cm}$ is input. It is now evident that for $B^{}_0 \approx 10^{24}~{\rm G}$ and $L^{}_0 > L^{\rm min}_0$, as predicted by a specific model in Ref.~\cite{Vachaspati:1991nm}, $\nu^{}_{\rm R}$ can be in thermal equilibrium at earlier times and then decouple from the thermal bath just before the QCD phase transition.

{\it Secret $\nu^{}_{\rm R}$ Interactions.}---It is also possible to populate $\nu^{}_{\rm R}$ via an exotic interaction, which is introduced exclusively for right-handed neutrinos and thus named as ``secret $\nu^{}_{\rm R}$ interaction". For our purpose, it is enough to follow a phenomenological approach and simply add two interaction terms into the SM
\begin{eqnarray} \label{eq:secret}
{\cal L} \supset - g^{}_\nu \overline{\nu^{}_{i{\rm R}}} \gamma^\mu \nu^{}_{i{\rm R}} V^{}_\mu - g^{}_\chi \overline{\chi} \gamma^\mu \chi V^{}_\mu \; ,
\end{eqnarray}
where $V^{}_\mu$ denotes a light vector boson $V$ (e.g., $m^{}_V \sim 1~{\rm MeV}$), while $\chi$ a heavy Dirac fermion (e.g., $m^{}_\chi \sim 2~{\rm TeV}$) as a candidate for cold dark matter. Here $g^{}_\nu$ and $g^{}_\chi$ stand for the coupling constants of $V$ with $\nu^{}_{\rm R}$ and $\chi$, respectively.
This kind of secret neutrino interaction was first considered for the SM left-handed neutrinos~\cite{Aarssen:2012fx,Ahlgren:2013wba,Laha:2013xua}, and later extended to sterile neutrinos of eV-scale masses~\cite{Dasgupta:2013zpn,Bringmann:2013vra,Ko:2014bka,Saviano:2014esa,Chu:2015ipa}. In these works, the secret interaction also applies to a dark matter particle, offering an intriguing solution to the small-scale problems of cold dark matter in the cosmological structure formation~\cite{Aarssen:2012fx}.

In this model, we further postulate that the dark matter sector $\chi$ is also coupled to the SM via some high-energy dynamics, such that $\chi$ can be thermally produced. However, the decoupling between these two sectors also takes place quite early, e.g., at $T^{}_{\rm d} \gg 10~{\rm TeV}$. Then, the new interactions in Eq.~(\ref{eq:secret}) will bring both $V$ and $\nu^{}_{\rm R}$ into thermal equilibrium with $\chi$, if the coupling constants are not extremely small. The evolution of the whole system can be summarized as follows.
\begin{itemize}
\item The dark matter annihilation $\chi \overline{\chi} \to V V$ and $\chi \overline{\chi} \to \nu^{}_{\rm R} \overline{\nu}^{}_{\rm L}$ (negligible for $g^{}_\nu \ll g^{}_\chi$) will be frozen out at $T^{}_\chi \approx m^{}_\chi/25$, leading to a correct relic dark matter density for $m^{}_\chi = 2~{\rm TeV}$ and $g^{}_\chi \approx 0.8$, as demonstrated in Ref.~\cite{Aarssen:2012fx}. The $\nu^{}_{\rm R}$-$\chi$ elastic scattering could keep them in kinetic equilibrium until $T \sim {\rm keV}$. This salient feature, together with the self-interaction of $\chi$ mediated by $V$, could help solve all the small-scale problems of structure formation.

\item Since $\nu^{}_{\rm R}$ is only in contact with the dark sector (i.e., $V$ and $\chi$), which decouples from the SM sector at $T^{}_{\rm d}$, it just cools down as $T^{}_{\nu^{}_{\rm R}} = [g^{}_{*s} (T^{}_\gamma)/g^{}_{*s}(T^{}_{\rm d})]^{1/3} T^{}_\gamma$, where $T^{}_\gamma$ represents the temperature of the SM sector. At the time of BBN, the extra number of neutrinos can be estimated as~\cite{Dasgupta:2013zpn}
    \begin{eqnarray}
    \Delta N^{}_{\rm eff} =  \frac{\rho^{}_{\nu^{}_{\rm R}} + \rho^{}_V}{(\rho^{}_{\nu^{}_{\rm L}}/3)} \approx \left[3 + \frac{3}{2} \times \frac{8}{7} \right] \times \left(\frac{10.75}{106.75}\right)^{4/3} \approx 0.22 \; ,
    \end{eqnarray}
    which is compatible with the BBN constraint. Later on, a reheating of $\nu^{}_{\rm R}$ and $\overline{\nu}^{}_{\rm L}$ arises from $V \to \nu^{}_{\rm R} \overline{\nu}^{}_{\rm L}$ when $V$ becomes non-relativistic at $T < m^{}_V$. The ratio of the $\nu^{}_{\rm R}$ temperatures after and before reheating is $T^\prime_{\nu^{}_{\rm R}}/T^{}_{\nu^{}_{\rm R}} = (11/7)^{1/3}$, whereas the photon temperature is increased by electron-positron annihilation, so the effective number of neutrinos in the epoch of CMB turns out to be
    \begin{eqnarray}
    \Delta N^{}_{\rm eff} = \frac{\rho^{}_{\nu^{}_{\rm R}}}{(\rho^{}_{\nu^{}_{\rm L}}/3)} \approx 3\times \left[ \frac{g^{}_{*s}(T^{}_{\rm CMB})}{g^{}_{*s}(T^{}_{\rm d})}\right]^{4/3} \times \left(\frac{11}{7}\right)^{4/3} \times \left(\frac{11}{4}\right)^{4/3} \approx 0.26 \; ,
    \end{eqnarray}
    where $g^{}_{*s}(T^{}_{\rm CMB}) = 3.91$ is used. Therefore, this model survives both BBN and CMB bounds. The number density of $\nu^{}_{\rm R}$ and $\overline{\nu}^{}_{\rm L}$ at present is $34 \times 11/7 \approx 53~{\rm cm}^{-3}$, as expected for a high freeze-out temperature and a late-time reheating from $V$ decays.
\end{itemize}

It is worth mentioning that an interesting scenario of non-thermal production of $\nu^{}_{\rm R}$ has been recently presented in Ref.~\cite{Chen:2015dka}, where the coupling between inflaton and $\nu^{}_{\rm R}$ is introduced to generate a degenerate $\nu^{}_{\rm R}$ gas via inflaton decays. After its production, $\nu^{}_{\rm R}$ evolves separately and is diluted as the Universe cools down. Although the energy density of $\nu^{}_{\rm R}$ is constrained by the BBN and CMB observations, as in our two scenarios, the relic number density of $\nu^{}_{\rm R}$ and $\overline{\nu}^{}_{\rm L}$ can be as high as one half of the photon density, i.e., around $220~{\rm cm}^{-3}$. The main reason is that $\nu^{}_{\rm R}$ is confined in the low-energy part of the non-thermal distribution function, compared to the thermal spectrum.

\section{Impact on the Detection of C$\nu$B}

Finally, we examine the impact of relic right-handed Dirac neutrinos on the detection of C$\nu$B in a future experiment, such as PTOLEMY~\cite{Betts:2013uya}. Experimentally, one studies the spectrum of the emitted electrons (e.g., from $\nu^{}_e + {^3{\rm H}} \to {^3{\rm He}} + e^-$), and looks for events that have kinetic energies above the beta-decay endpoint. However, because of tiny neutrino masses and small momenta carried by relic neutrinos, the expected signals are very close to the endpoint. For instance, if three active neutrinos have a degenerate mass $m^{}_i \approx m_0^{}$, we then expect the signal to show up at the position that is $2m_0^{}$ beyond the endpoint. An energy resolution comparable to the neutrino mass $m_0^{}$ is then required in order to clearly select the signal from the dominant beta-decay background. In the recent proposal PTOLEMY, new techniques are suggested to achieve such a high energy resolution. The beta-decaying nuclei ${^3}{\rm H}$ will be deposited onto some surface substrate so as to reduce the nucleus recoil. An energy resolution $\Delta \sim 0.15~\rm{eV}$, defined as the full width at half maximum (FWHM) of the Gaussian distribution, is claimed to be achievable, so the detection of C$\nu$B is promising, in particular for nearly-degenerate neutrino masses and relatively large values of absolute neutrino masses (e.g., $m^{}_0 > \Delta$).

Beside the location of the signal, one can also measure its height, which is related to the capture rate. Adopting the calculation from~\cite{Long:2014zva}, we find the capture rate of relic neutrinos
\begin{eqnarray}
\Gamma_{\rm C \nu B}^{} = \sum_{s_\nu^{} =\pm 1/2} ~\sum_{j=1}^{3}~ \sigma_j^{} (s_\nu^{})  v_{\nu_j^{}}^{}  n_j^{} (s_\nu^{}) N_T^{} \; ,
\end{eqnarray}
where $s_\nu^{}$ stands for the two helical states, and $j$ indicates three different mass eigenstates. In addition, $n_j^{}(s_\nu^{})$ represents the number density of incoming relic neutrino $\nu_j^{}$ in the helical state $s_\nu^{}$, while $N_T^{} = M_T^{}/m_{\rm ^{3}H}$ is the number of target nuclei. Approximately, we have $\sigma_j^{} (s_\nu^{})  v_{\nu_j^{}}^{} \simeq A(s_\nu^{}) |U_{ej}^{}|^2 \bar{\sigma}$, where $U_{ej}^{}$ (for $j = 1, 2, 3$) are the matrix elements in the first row of the lepton mixing matrix, and $\bar{\sigma} \simeq 3.834 \times 10^{-45} ~{\rm cm}^2$ for the neutrino capture on tritium. More importantly, the spin-dependent factor $A(s_\nu^{})$ is given by~\cite{Long:2014zva}
\begin{eqnarray}
A(s_\nu^{}) \equiv 1 - 2 s_\nu^{} v_{\nu_j^{}} =
\begin{cases}
1 - v_{\nu_j}, \quad s_\nu^{} = +1/2 \quad {\rm right~helical} \\
1 + v_{\nu_j}, \quad s_\nu^{} = -1/2 \quad {\rm left~helical} \; ,
\end{cases}
\end{eqnarray}
where the neutrino velocity $v_{\nu_j^{}} = |\mathbf{p}_{\nu_j^{}}^{}|/E_{\nu_j^{}}^{}$ approximates to 0 and 1 in the non-relativistic and relativistic limits, respectively. According to the standard cosmology, the present temperature of ${\rm C \nu B}$ is around $T_\nu^{} = 0.168~{\rm meV}$. Therefore, for the degenerate mass region that can be probed by PTOLEMY, relic neutrinos are highly non-relativistic today. In this case, we have $A(\pm 1/2) = 1$ for both left and right helical states. As shown in Eq.~(\ref{eq:rates}), if both $\nu^{}_{\rm R}$ and $\overline{\nu}^{}_{\rm L}$ are absent in the cosmic background, we have the capture rate $\Gamma^{}_{\rm D} \approx 4~{\rm yr}^{-1}$ for 100 grams of tritium. This rate will be modified in the following cases.
\begin{enumerate}
\item If both $\nu^{}_{\rm R}$ and $\overline{\nu}^{}_{\rm L}$ are thermally produced and decouple from the thermal bath at a high temperature above the electroweak phase transition. The number density of right- and left-helical neutrino states will be increased by $10\%$, implying a capture rate $\Gamma^{\rm R}_{\rm D} \approx 4.4~{\rm yr}^{-1}$. However, if they freeze out just before the QCD phase transition, the capture rate can be enhanced by $28\%$, namely, $\Gamma^{\rm R}_{\rm D} \approx 5.1~{\rm yr}^{-1}$. The latter scenario saturates the upper bound on the extra effective number of neutrinos in the BBN and CMB eras.

\item Unlike the thermal production, the non-thermal and degenerate gas of $\nu^{}_{\rm R}$ and $\overline{\nu}^{}_{\rm L}$ considered in Ref.~\cite{Chen:2015dka} could change the capture rate by $64\%$, namely, $\Gamma^{\rm R}_{\rm D} \approx 6.6~{\rm yr}^{-1}$. This modification will diminish the chance to distinguish between Dirac and Majorana neutrinos via detection of C$\nu$B.
\end{enumerate}

To illustrate the impact of the existence of relic right-handed neutrinos, we draw the expected spectrum of electrons by assuming that the background arises only from the beta decays of tritium. Moreover, to account for the effects due to a finite energy resolution, we follow the approach given in Ref.~\cite{Long:2014zva} by convolving both the beta-decay spectrum and the true C$\nu$B signal with a Gaussian envelope of FWHM $\Delta$. Under these assumptions, we consider two benchmark scenarios by taking $(\Delta, m_0^{})$ as $(0.15~{\rm eV}, 0.25~{\rm eV})$ and $(0.05~{\rm eV}, 0.07~{\rm eV})$, respectively. These two sets of benchmark values are chosen to respect the rule of $\Delta \lesssim 0.7 m_0^{}$, which was found to be the necessary condition of discovering relic neutrinos from beta-decay background \cite{Long:2014zva}. In addition, for the first benchmark scenario we choose its energy resolution to be compatible with the PTOLEMY experiment, while in the second one we reduce the energy resolution so as to meet the neutrino mass requirement $\Sigma < 0.23~{\rm eV}$ at the $95\%$ C.L. from the latest Planck result \cite{Ade:2015xua}.

\begin{figure}[t!]
\begin{center}
\subfigure{
\includegraphics[width=0.48\textwidth]{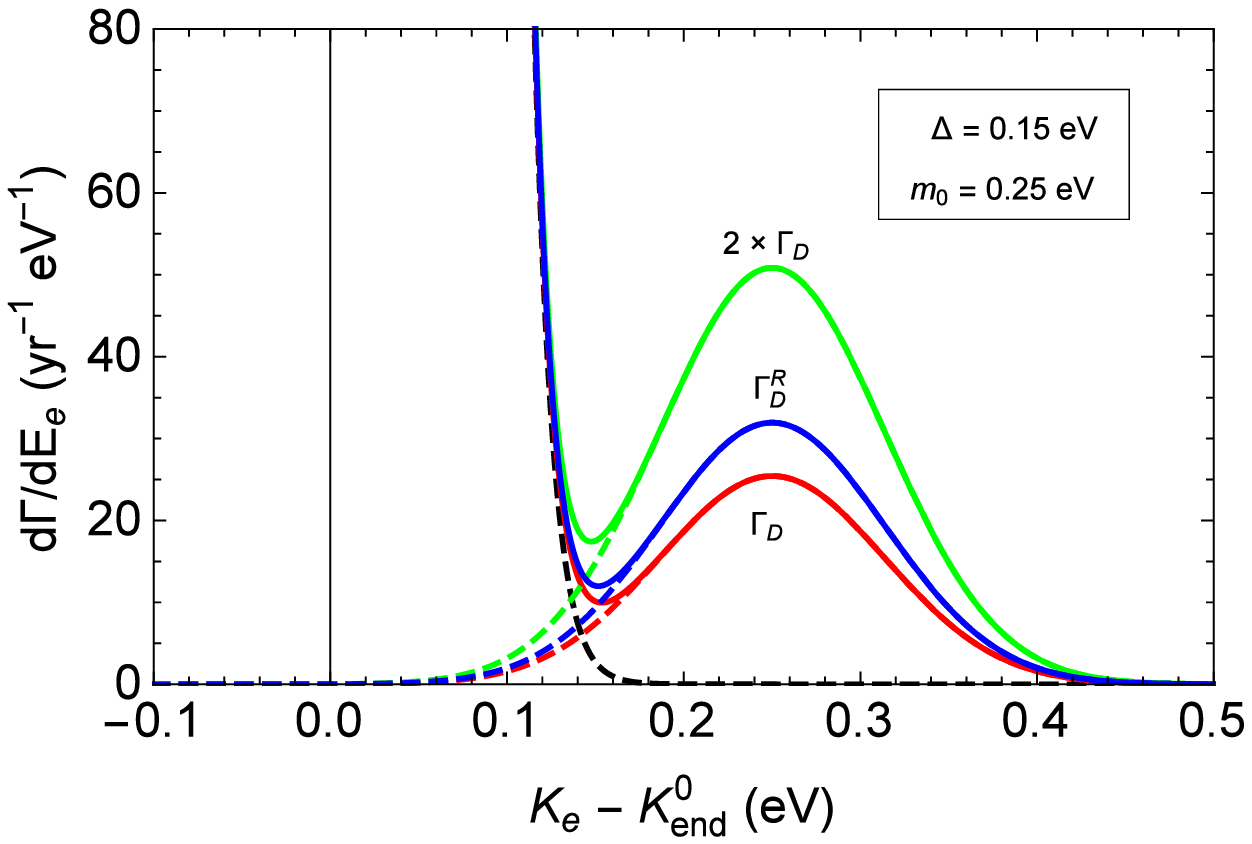} }
\subfigure{
\includegraphics[width=0.48\textwidth]{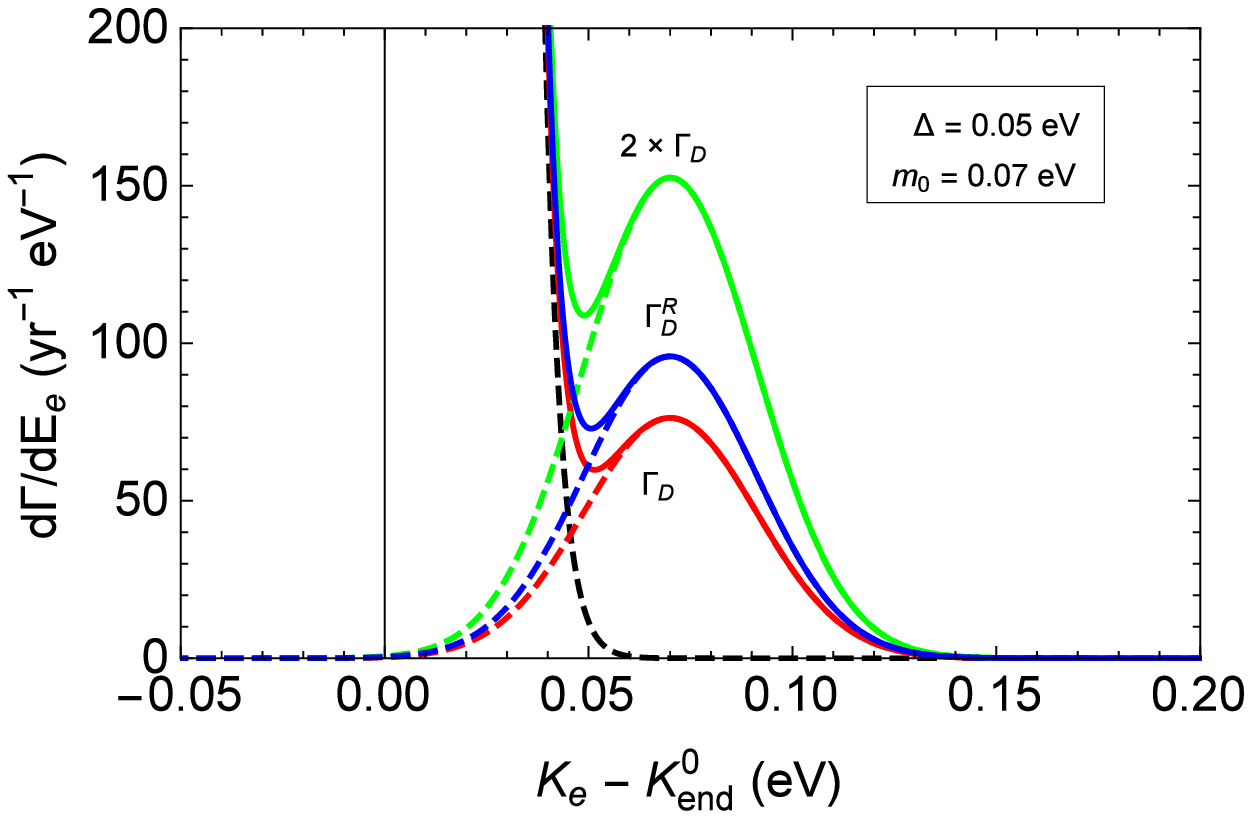} }
\vspace{-0.6cm}
\caption{Expected spectra of emitted electrons in PTOLEMY with two sets of energy resolution $\Delta$ and neutrino mass $m_0^{}$. The black dashed curves are the background contribution from beta decays, while the colored dashed curves are the possible signals with three different capture rates assumed, i.e., red for the standard Dirac case $\Gamma_{\rm D}^{} = 4 ~{\rm yr}^{-1}$, blue for the case with the inclusion of right-handed neutrinos $\Gamma_{\rm D}^{\rm R} = 5.1 ~{\rm yr}^{-1}$, and green for that twice the standard Dirac contribution. The observed spectra are denoted by colored solid curves, which are the sum of the background and signal contributions.
}
\label{fig:PTOLEMY_spectrum}
\end{center}
\end{figure}

The expected spectra of electrons for the above benchmark scenarios are drawn in Fig.~\ref{fig:PTOLEMY_spectrum}, given different values of the total capture rate. Here, the differential capture rate ${\rm d}\Gamma/{\rm d} E_{\rm e}$ has been shown as a function of the kinetic energy of electrons $K_{\rm e}^{}$ calibrated by $K_{\rm end}^0$, which is the beta-decay endpoint in the limit of massless neutrinos. The standard scenario of massive Dirac neutrinos with $\Gamma_{\rm D}^{} = 4~{\rm yr}^{-1}$ is represented by the red solid curve, while the new-physics scenario with $\Gamma_{\rm D}^{\rm R} = 5.1~{\rm yr}^{-1}$ by a blue solid curve. For comparison, the capture rate $\Gamma^{}_{\rm M} = 2 \Gamma^{}_{\rm D}$ in the case of Majorana neutrinos has also been given as a green solid curve. As one can see, because of the enhancement of the capture rate, the observed spectrum in the latter case has a peak higher than the former one. Therefore, if one can experimentally resolve such a difference in the spectrum, a discrimination between the standard Dirac scenario and its modification might be possible. We next study such a discrimination quantitatively by taking the first benchmark scenario as an example, as it more closely resembles the proposed PTOLEMY experiment.
%%%%%%%%%%%%%%%%%%%%%%%%%% Fig. 2 %%%%%%%%%%%%%%%%%%%%%%%%%%%%
%%%%%%%%%%%%%%%%%%%%%%%%%%%%%%%%%%%%%%%%%%%%%%%%%%%%%%%%%%%%%%
First, we take the capture rate in the standard Dirac case (i.e., $\Gamma = 4 {\rm yr}^{-1}$) as the true signal rate, and generate the true spectrum of electrons for a specific choice of data-taking period. Then, we fit this true spectrum by two parameters, namely, the capture rate $\Gamma$ and neutrino mass $m_0^{}$. In the fitting, we consider a region of interest that spans from the zero of $K_{\rm e}^{} - K_{\rm end}^{0}$ and towards the signal end, and take as many bins as possible to sufficiently cover the signal region, where a bin size of $0.15~{\rm eV}$ identical to the energy resolution has been chosen. Therefore, for the first benchmark scenario, the energy region of interest is $[0, ~0.45]~{\rm eV}$, divided into three bins. It is worthwhile to point out that this method is independent of the location of signal peak (i.e., the neutrino mass). The number of events in each bin is then used to calculate the probability distribution of the fitted capture rate $\Gamma$ and neutrino mass $m_0^{}$, by assuming a Poisson distribution of the event number.

\begin{figure}[t!]
\begin{center}
\subfigure{
\includegraphics[width=0.55\textwidth]{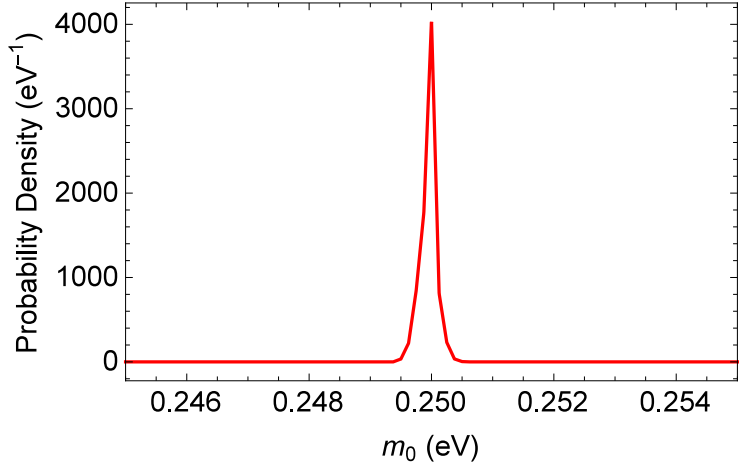} }
\vspace{-0.1cm}
\caption{The probability distribution of the neutrino mass $m^{}_0$, where the true value $m^{}_0 = 0.25~{\rm eV}$ and one year of data have been assumed and the capture rate $\Gamma$ has been marginalized over. }
\label{fig:PTOLEMY_m0}
\end{center}
\end{figure}

By carefully inspecting the two-dimensional probability distribution of $\Gamma$ and $m_0^{}$, we find that the neutrino mass can be very precisely determined. In Fig. \ref{fig:PTOLEMY_m0}, we plot the normalized probability distribution of the neutrino mass $m_0^{}$ by marginalizing over the capture rate $\Gamma$. As one can see, even with one year of data-taking, the fitted $m_0^{}$ is well peaked at the true value of $m_0^{} = 0.25 ~{\rm eV}$ and its width is extremely small. This is due to the fact that if the fitted neutrino mass slightly differs from the true value, the corresponding spectrum, especially the beta decay part, would have a drastic distortion, resulting in a good discriminating power for the neutrino mass. However, one should also keep in mind that in reality there might be large systematic uncertainties in modeling the beta-decay spectrum, such as the uncertainty of the energy resolution.

\begin{figure}[t!]
\begin{center}
\subfigure{
\includegraphics[width=0.6\textwidth]{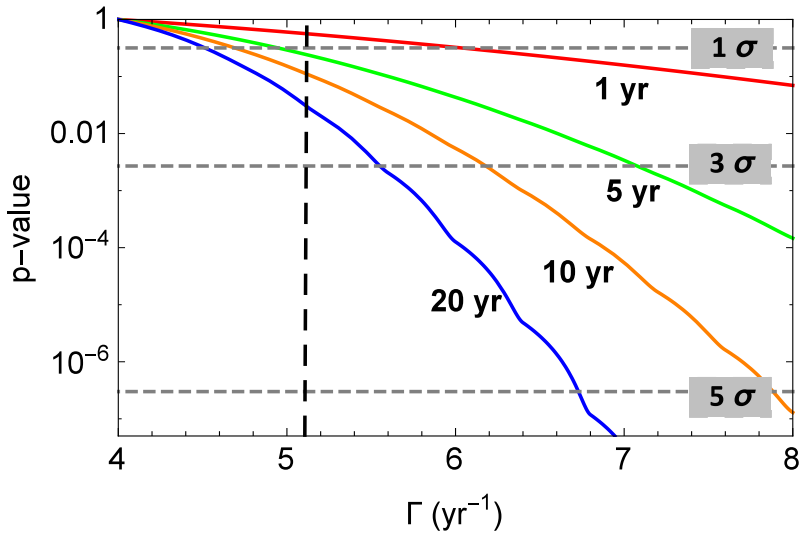} }
\vspace{-0.1cm}
\caption{The $p$-value is shown for the test value of the capture rate $\Gamma$, where the true signal has been assumed to be $\Gamma_{\rm D} = 4~{\rm yr}^{-1}$ in the standard Dirac case. The energy resolution $\Delta$ and the true value of neutrino mass $m_0^{}$ are chosen to be $0.15~{\rm eV}$ and $0.25~{\rm eV}$, respectively. The vertical dashed line is for the reference value $\Gamma_{\rm D}^{\rm R} = 5.1~{\rm yr}^{-1}$.
}
\label{fig:PTOLEMY_sigma}
\end{center}
\end{figure}

Having reconstructed the neutrino mass from the spectrum of electrons, we now examine the discrimination among different capture rates. To do so, we first derive the probability distribution of the capture rate $\Gamma$ by marginalizing over the neutrino mass $m_0^{}$. Then, for a given capture rate $\Gamma$, the $p$-value is calculated as the exclusion probability by integrating the distribution function over the values larger than $\Gamma$~\cite{Cowan:2010js}. In Fig.~\ref{fig:PTOLEMY_sigma} we show the $p$-value as a function of the capture rate. Several different data-taking periods are considered. It can be seen that for the non-standard Dirac scenario with $\Gamma_{\rm D}^{\rm R} = 5.1 ~{\rm yr}^{-1}$ (the vertical dashed line), running PTOLEMY-like experiments for about 5 years is only able to distinguish it from the standard one at $1\sigma$ level, and to reach a $3\sigma$ exclusion a detection time of even 20 years is not enough. Therefore, one may consider increasing the target mass in order to reduce the required data-taking time.  As for the discrimination between the Dirac and Majorana scenarios, from Fig.~\ref{fig:PTOLEMY_sigma}, we observe that a data-taking time of 5 years is sufficient for distinguishing these two scenarios at $3\sigma$ level, while a detection time of about 10 years is required to achieve a $5\sigma$ significance.
%%%%%%%%%%%%%%%%%%%%%%%%%% Fig. 3 %%%%%%%%%%%%%%%%%%%%%%%%%%%%
%%%%%%%%%%%%%%%%%%%%%%%%%% Fig. 4 %%%%%%%%%%%%%%%%%%%%%%%%%%%%
%%%%%%%%%%%%%%%%%%%%%%%%%%%%%%%%%%%%%%%%%%%%%%%%%%%%%%%%%%%%%%

Although our statistical analysis is performed for a particular choice of energy resolution and neutrino mass, it may apply to other similar scenarios, where the signal spectrum is well separated from the beta-decay background (i.e., there exist energy bins in which signal contributions are dominant). However, for the cases where the signal and background are not easy to separate, one would need a longer data-taking time. In this sense, the results given in Fig.~\ref{fig:PTOLEMY_sigma} may be regarded as the most optimistic case, and the indicated detection time can be viewed as lower bounds for more general choices of energy resolution and neutrino mass.

In the above discussions, we have used the average number densities of neutrinos to calculate the capture rates of C$\nu$B, but it should be noticed that an overabundance of neutrinos within the dark matter halo is possible via the gravitational clustering~\cite{Ringwald:2004np,Long:2014zva}. Depending on the absolute neutrino masses, the clustering effects could be significant. For the absolute scale of neutrino masses $m^{}_0 = 0.15~{\rm eV}$, the capture rate will be enhanced by a factor of $1.4$ if the Navarro-Frenk-White profile~\cite{Navarro:1996gj} of the dark matter in our galaxy is assumed, or by a factor of $1.6$ for the Milky Way model~\cite{Klypin:2001xu}. Hence, the unknown dark matter profile leads to the remarkable uncertainty of gravitational clustering, which can be even larger for heavier neutrinos, rendering the discrimination between Dirac and Majorana neutrinos extremely challenging. Additionally, in the presence of relic right-handed neutrinos, the difference between the capture rates $\Gamma^{}_{\rm D}$ and $\Gamma^{}_{\rm M}$ becomes even smaller. Therefore, the uncertainty in gravitational clustering effects and a possible cosmic background of right-handed neutrinos will diminish the experimental discriminating power on the nature of neutrinos.

Finally, it is also interesting to think about a possible way to discriminate the thermal and non-thermal production mechanisms of right-handed Dirac neutrinos. The neutrino energy spectra are quite different in these two cases. For instance, the non-thermal distribution proposed in Ref.~\cite{Chen:2015dka} favors low-energy neutrino states, implying the importance of measuring the velocities of final-state electrons. However, the terms involving the incident neutrino momenta in the capture rate are neglected, as they are typically very small and further suppressed by the heavy mass of the target nuclei. Hence, the final capture rate $\Gamma_{\rm C\nu B}^{}$ mainly depends on the number density of the incoming relic neutrinos, while the velocity distribution or the detailed energy spectrum of relic neutrinos is almost irrelevant.

\section{Summary}

In this paper, assuming massive Dirac neutrinos, we have considered the impact of right-handed neutrinos on the detection of cosmic neutrino background in the future PTOLEMY experiment, in which the beta-decaying tritium will be used to capture background neutrinos.

First, we demonstrate that the production rate of right-handed neutrinos is extremely small in the extended SM, as already found in Ref.~\cite{Antonelli:1981eg}. Although our calculations show that the rate can actually be larger by several orders of magnitude than that in Ref.~\cite{Antonelli:1981eg}, it is still insufficient for right-handed neutrinos to be thermally produced. Second, in assumption of thermal right-handed neutrinos present in the early Universe, we find that the cosmological constraints on the effective number of neutrinos require them to decouple from thermal bath at latest in the epoch of QCD phase transition. When the cosmological upper bound $\Delta N^{}_{\rm eff} \lesssim 0.53$ is saturated, we obtain the right-to-left ratio of neutrino number densities $n^{}_{\nu_{\rm R}}/n^{}_{\nu_{\rm L}} \approx 0.28$. Namely, the relic density of right-handed neutrinos $\nu^{}_{\rm R}$ and left-handed antineutrinos $\overline{\nu}^{}_{\rm L}$ can be as large as $95~{\rm cm}^{-1}$. Third, we present two possible scenarios to realize thermal production of right-handed Dirac neutrinos. The first one is just to utilize the magnetic dipole moments of massive Dirac neutrinos in the SM, given primordial magnetic fields generated in the electroweak phase transition. The second one is to introduce secret interactions among right-handed neutrinos and cold dark matter, which could help solve the small-scale structure problems. Finally, we examine how the presence of right-handed neutrinos affects the capture rate of cosmic neutrino background. Quantitatively, for massive Dirac neutrinos, the capture rate can be enhanced from $\Gamma^{}_{\rm D} = 4~{\rm yr}^{-1}$ to $5.1~{\rm yr}^{-1}$ in the PTOLEMY experiment with 100 grams of tritium~\cite{Long:2014zva,Betts:2013uya}. To observe the impact of right-handed neutrinos, a data-taking time about 5 years is needed to reach a statistical significance of $1\sigma$ in the most optimistic case. Therefore, it seems difficult to exclude or prove the existence of relic right-handed neutrinos in the near future.

The proposal of PTOLEMY experiment and its great physics potential have stimulated us to take more seriously the detection of cosmic neutrino background and some related issues, such as the presence of right-handed Dirac neutrinos and the discrimination between Dirac and Majorana neutrinos. Further progress in both experimental and theoretical studies in this direction will hopefully extend greatly our knowledge on the intrinsic properties of neutrinos.

\hspace{0.5cm}
\begin{flushleft}
{\bf Note Added:}
\end{flushleft}
When the present paper was in preparation, Ref.~\cite{Chen:2015dka} appeared in the preprint archive. Although both our paper and Ref.~\cite{Chen:2015dka} have considered the impact of relic right-handed neutrinos on the detection of cosmic neutrino background, the production mechanisms for right-handed neutrinos are different and complementary. After finishing this work, we became aware of Refs.~\cite{Anchordoqui:2011nh,Anchordoqui:2012qu,SolagurenBeascoa:2012cz}, where the right-handed neutrinos as dark radiation have been discussed in a $U(1)$ gauge model with a heavy $Z^\prime$ gauge boson.

\hspace{0.5cm}
\begin{flushleft}
{\bf Acknowledgements}
\end{flushleft}
The authors are indebted to Prof. Xiangdong Ji for an inspiring question and helpful discussions during Symposium on Physics Potential of China Jinping Underground Laboratory in May, 2015 at Tsinghua University, Beijing, and to Prof. Zhi-zhong Xing for his great interest in this work and encouragements. This work was supported in part by the Innovation Program of the Institute of High Energy Physics under Grant No. Y4515570U1, by the
National Youth Thousand Talents Program, and by the CAS Center for
Excellence in Particle Physics (CCEPP).

\end{document}